# Experimental Evidence for Coulomb Charging Effects in the Submicron Bi-2212 Stacks

Yu. I. Latyshev,[1-3]  S.-J. Kim,[2,3] and T. Yamashita[2,3]

[1] *Institute of Radio-Engineering and Electronics Russian Acad. of Sciences, 11 Mokhovaya str., 103907 Moscow, Russia*

[2] *Research Institute of Electrical Communication, Tohoku University, 2-1-1, Katahira, Aoba-ku, Sendai 980-8577, Japan*

[3] *CREST, Japan Science and Technology Corporation (JST), Japan*

We developed the focused ion beam (FIB) and ion milling techniques for a fabrication of the $Bi_2Sr_2CaCu_2O_{8+\delta}$ (*Bi-2212*) stacked junctions with in-plane size $L_{ab}$ from several microns down to the submicron scale without degradation of $T_c$. We found that behaviour of submicron junctions ($L_{ab} < 1$ μm) is quite different from the bigger ones. The critical current density is considerably suppressed, the hysteresis and multibranched structure of the *IV* characteristics are eliminated, the periodic structure of current peaks reproducibly appears on the *IV* curves at low temperatures. A period of the structure, $\Delta V$, is consistent with the Coulomb charging energy of a single pair, $\Delta V = e/C$ with *C* the effective capacitance of the stack. We consider this behaviour to originate from the Coulomb blockade of the intrinsic Josephson tunneling in submicron *Bi-2212* stacks.

As it is well known, the tunneling current in a tunnel junction of small capacitance $C$ can be blocked by the charging effect at the bias voltages, $eV$, below the Coulomb charging energy of a single electron, $E_c = e^2/2C$, when $E_c$ exceeds the energy of thermal and quantum fluctuations, $E_c > kT$, $R > h/4e^2 = R_Q$ with $R$ the tunneling resistance of the junction [1]. In a similar way the Coulomb blockade effect can block the Josephson tunneling current, in a Josephson junction of small enough area at the low temperature when $E_c$ becomes comparable or exceeds the Josephson coupling energy $E_J$ [2]. The Coulomb blockade of the Cooper pair tunneling leads to an appearance on the current-voltage (*IV*) characteristics of small junctions the voltage periodic structure of current peaks with a period $\Delta V = 2E_c$ [3]. The "supercurrent" has a finite slope due to the quantum or classical diffusion of the phase of the order parameter [2] and the value of supercurrent is suppressed proportionally to the ratio $E_c / E_J$ [2].

The charging effects become even more strong in the arrays of the small junctions [4], where the tunneling of a single electron (or pair) has been recently considered as the correlated motion of the charge soliton including a number of junctions [5,6]. The charge soliton length can be enormously large in the vertically stacked junctions, because of their negligibly small stray capacitance [6]. It implies that the whole stack with $N$ junctions can effectively work as a single unit with the charge energy being $N$ times higher than the charging energy of a single junction.

In this paper we report on studies of the intrinsic Josephson effect [7] in the stacked structures of *Bi-2212* with a successive decrease of their in-plane area, $S$, down to the submicron scale. We found that the interplay between the charging effects and the intrinsic Josephson effect becomes remarkable at the $S \sim 1$ μm$^2$. For the submicron junctions we clearly observed the periodic structure of current peaks on the *IV* characteristics with a period corresponding to the charge energy of the Cooper pair charge soliton, including the total number of the elementary junctions ( ~50) of the stack.

We used the $Bi_2Sr_2CaCu_2O_{8+\delta}$ (*Bi-2212*) whiskers [8] as the base material for the stacks fabrication. Whiskers have been grown by the *Pb*-free method [9] and have been characterized by TEM as a very perfect crystalline object [9]. We developed the FIB and the ion milling techniques for a fabrication of the *Bi-2212* stacked junctions with in-plane size from several microns down to the submicron scale without degradation of $T_c$ [10]. The stages of the fabrication are shown at Fig. 1. We used for fabrication the conventional FIB machine of Seiko Instruments Corp., SMI-900 (SP) operating with the $Ga^+$-ion beam with the energy ranging from 15 to 30 keV and the beam current from 8 pA to 50 nA. For the smallest current the beam diameter can be focused down to 10 nm. We estimate the maximum penetration depth of 30 kV $Ga^+$ ions along the *c*-axis in *BSCCO* single crystal to be 60 nm and the lateral scattering depth to be 40 nm [11]. For the ion milling we used standard *Ar*-ion plasma technique. Four electrical contacts have been prepared beyond the zone of the processing in two ways, by silver paste or by evaporation of 70 nm-thick golden pads with the following annealing in the oxygen flow at 450°C. In both cases the contact annealing has been done before the FIB processing to avoid a diffusion of the *Ga*-ions in the stack body. The parameters of several stacks under investigation are listed in the Table.

The *IV* characteristics have been measured in a shielded room using the low noise preamplifier and the high sensitive oscilloscope at frequencies ~ 100 Hz. We estimated the noise current level to be about 10 nA. The temperature dependences of the c-axis resistivity $\rho_c$ have been measured using *DC*-currents of 1 - 5 μA.

The $\rho_c(T)$ dependences of the stacks were typical to the slightly overdoped *Bi-2212* case [12], with $T_c \approx 77$ K and $\rho_c (300) \approx 10\text{-}12$ $\Omega$ cm. The critical current density along the *c*-axis for the junctions with in-plane area $S > 2$ $\mu m^2$ was $J_c \approx 6\ 10^2$ A/cm$^2$ which is consistent with a theoretical estimation, $J_c \approx 8\ 10^2$ A/cm$^2$, given by [13]: $J_c(0) = c\Phi_0/(8\pi^2 s\lambda_c^2)$, with $\Phi_0$ the flux quantum, $\lambda_{ab} = 0.2$ $\mu$m and $\gamma = \lambda_c / \lambda_{ab} \approx 1000$ [14]. The dependences of critical current along the *c*-axis on parallel magnetic field for bigger stacks demonstrate quite good Fraunhoffer patterns [15] which proved the presence of the DC intrinsic Josephson effect in our stacks.

Fig.2 shows the large current and voltage scale *IV* characteristics of three stacks with a decrease of $S$ down to 0.3 $\mu m^2$. The gap voltage at $V = V_g$ and the normal state resistance $R_N$ at $V > V_g$ are well defined. The $R_N$ was practically temperature independent and corresponds to the resistivity value $\rho_c (V > V_g) \approx 12$ $\Omega$ cm. It gives a possibility to estimate the number of the elementary junctions, $N$, using a simple formula: $N = R_N S / \rho_c(V > V_g) s$. For the submicron junctions (Figs. 2b, c) we have not observed the *S*-shaped *IV* characteristics. It implies that the self-heating [16] and the non-equilibrium injection effects [17] are essentially eliminated. As a result we found that the superconducting gap of the elementary junction $2 \Delta_0 = e V_g / N$ reaches in submicron junctions a value $2 \Delta_0 \approx 50$ meV (see Table), which is consistent with the value found recently from the surface tunneling measurements [18]. The value of $T_c$ determined by polynomial extrapolation of $\Delta(T)$ to zero lies within 76-78 K for all the samples including the smallest one with $S = 0.3$ $\mu m^2$ (see Table). It indicates that the junction body is not affected essentially by the FIB processing.

The *IV* characteristics in the extended scale are shown in the Fig. 3. For a bigger junction # 2 the multibranched structure [7] is clearly seen. A variation of the critical current along the stack is not big, indicating a good uniformity of our structures. The number of branches counted directly from the *IV* characteristics is consistent with the number calculated using the formula for $N$ given above.

We found also that with a decrease of $S$ below 1-2 μm$^2$ the critical current density of the stacks $J_c$ falls down (see Table). For the submicron stacks the multibranched structure disappears. The critical current transforms into a "supercurrent" peak located at finite voltage. Besides, a periodic set of current peaks appears (Fig.3). All these features are typical for the manifestation of the Coulomb blockade in small Josephson junction [3,19]. The period of the current peak structure, $\Delta V$, usually corresponds to the charging energy of a single pair, $2E_c$, [3]. For our smallest stack # 6 the period corresponds to $\Delta V = 1.1$ mV, i.e. about thrice the energy of thermal fluctuations, $kT$, at 4.2 K, and about 5 times higher than the Josephson coupling energy, $E_J = h\, I_c / (4\pi e)$, at this temperature.

In the stacked junction the charging effects are associated with the charge soliton [4-6]. The energy of the charge soliton, $E_s$, is proportional to the number of junctions located within the soliton length. The obtained big charging energy, $\Delta V \approx 1$ mV, *($\Delta V \equiv E_s$)* evidently corresponds to the number of junctions. The estimation of the charging energy of a single elementary junction shows that this number corresponds to the total number of the elementary junctions in the stack. For instance, for the stack #5 the charging energy of an elementary junction, $2E_{c0}$, estimated as $2E_{c0} = e^2 s / (\varepsilon_0 \varepsilon_c S)$ with $\varepsilon_c$ the dielectric constant along the *c*-axis, $\varepsilon_c = 5$ for *Bi-2212* [20], is $2E_{c0} \approx 20$ μeV and the ratio $e\, \Delta V / 2E_{c0}$ exactly corresponds to the $N = 50$, the total number of the junctions in this stack. We reproducibly observed the periodic structure of current peaks on 3 submicron stacks, with $R_N \geq 10$ k$\Omega$, in each case the period corresponding to the charge energy of the whole stack (see Table). That observation is consistent with an estimation of the soliton length [4]: $L_s = 2(C_0/C_g)^{1/2}$ with $C_g$ the stray capacitance, $C_g \approx (\varepsilon + 1) s /8$ [6], where $\varepsilon$ is the dielectric constant of the substrate. It gives for $L_s/s$ the value about 400. From this estimation it follows that the stack with $N < 400$ will respond to the transfer of a single pair as a single unit. It may be the reason of a disappearance of the multibranched structure in the submicron stacks.

With a decrease of $S$ the charging energy, $\Delta V$, should increase inversely proportional to the stack capacitance, $C$, or should be directly proportional to $R_N$. The latter statement results from the fact that the $R_N C$ product is a constant value, independent on the stack geometry and size: $R_N C = \rho_c \varepsilon_0 \varepsilon_c$. That is in a reasonable agreement with experiment. We found $\Delta V$ to be roughly proportional to $R_N$ for the three studied submicron stacks (see Table).

With an increase of the temperature above 4.2 $K$ the structure of peaks gradually washes out (Fig. 4) and disappears above 12 $K$ (for the stack # 6) when the charge soliton energy $\Delta V$ becomes less than $kT$. The behaviour of the supercurrent peak under conditions $E_J << kT$, $E_J << E_c$, which are roughly valid for our case, has been analysed in [21]. It was shown that in the case, when the resistance of the environment $Z_1$ is less than the quantum resistance $R_Q$, the junction will have classical phase diffusion behaviour and the zero bias resistance, $R_0$, should be proportional to $(kT)^2$: $R_0 = 2Z_1 (kT/E_J)^2$. That is in a qualitative agreement with our experiment (see insert to Fig.4).

A downfall of $J_c$ of submicron junctions can also be explained by the Coulomb blockade effect. According to Ref. [2], $J_c$ should drop down $\propto E_J/4E_c$. For our stacks the condition $E_J < 4 E_c$ met for $S < 1$ μm$^2$. We defined the critical current for the submicron junctions as a height of the supercurrent peak, $I_0$. The other current peaks, $I_n$, at $V = n\,\Delta V$ (n integer) also originate from the supercurrent contribution [3], corresponding in our case to the correlated motion of the single Cooper pair solitons. We can note here that the amplitude of the both types of peaks changes with a decrease of $S$ in the correlated way, i.e. $I_1/I_0 \approx const$ (compare Figs. 3b and 3c), indicating their common nature. The detailed picture of the correlated soliton motion in the stacked structures is still not clear and needs in the further theoretical and experimental investigations.

In summary, we developed the FIB method combined with the ion milling technique for a fabrication of the small *Bi-2212* stacked junctions with in-plane size down to the submicron level without degradation of their $T_c$. We found that a low temperature behaviour of the submicron junctions is governed by the Coulomb charging effects and their interplay with the intrinsic Josephson tunneling. That is the first observation of the Coulomb charging effects in the layered high-*Tc* materials.

The authors are acknowledged to K. K. Likharev, T. Claeson and L.N.Boulaevskii for a helpful discussion the results. The work has been partly supported by the Russian State Programme on HTSC (grant N 95028).

- e-mail: latychev@riec.tohoku.ac.jp

Table 1. Parameters of the stacked Bi-2212 junctions.

| No | S (μm$^2$) | $R_N$ (kΩ) | $I_c$ (μA) | $J_c$ (A/cm$^2$) | $T_c$ (K) | $V_g$ (V) | N | $V_g/N$ (mV) | C (fF) | ΔV (mV) | $2NE_{c0}$ (mV) |
|----|------|------|------|------|-----|-----|----|-----|------|------|------|
| #1 | 6.0  | 2.1  | 36   | 600  | 76  | 1.1 | 69 | 16  | 2.4  | -    | 0.67 |
| #2 | 2.0  | 6.5  | 12   | 600  | 76  | 1.3 | 65 | 20  | 0.68 | -    | 0.22 |
| #3 | 1.5  | 4.6  | 6    | 400  | 78  | 1.1 | 38 | 29  | 0.96 | -    | 0.18 |
| #4 | 0.6  | 10   | 0.24 | 40   | 76  | 1.7 | 34 | 50  | 0.44 | 0.38 | 0.36 |
| #5 | <1   | 13   | 0.30 | 30-60| -   | -   | -  | -   | 0.38 | 0.45 | 0.42 |
| #6 | 0.3  | 30   | 0.07 | 23   | 78  | 2.2 | 50 | 44  | 0.15 | 1.10 | 1.04 |

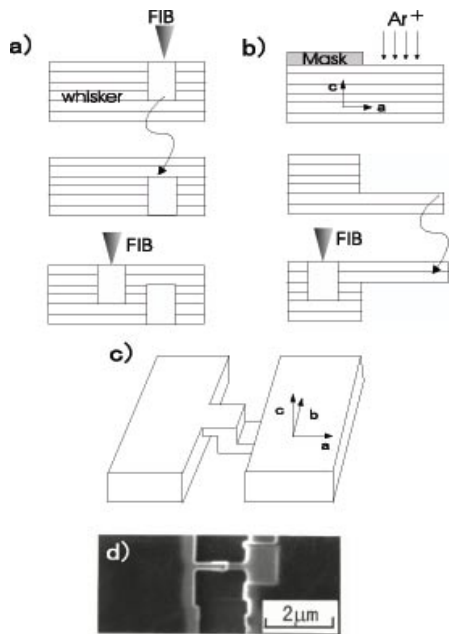

Fig.1. Stages of the stack fabrication with FIB (a), FIB combined with the ion milling (b), a schematic view (c), and a micrograph of the submicron *Bi-2212* stacked junction.

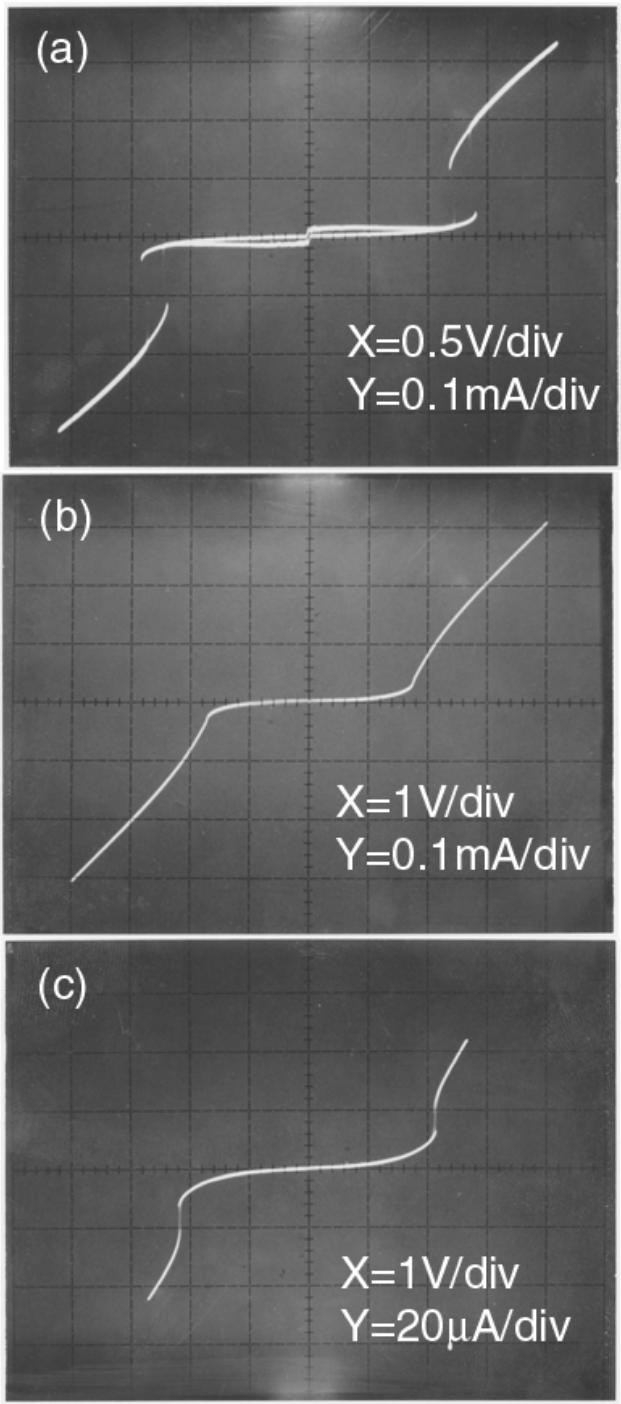

Fig. 2. The large current and voltage scale *I-V* characteristics of the *Bi-2212* stacks: (a) #2, $S = 2$ μm$^2$; (b) #4, $S = 0.6$ μm$^2$; (c) #6, $S = 0.3$ μm$^2$. T = 4.2 K.

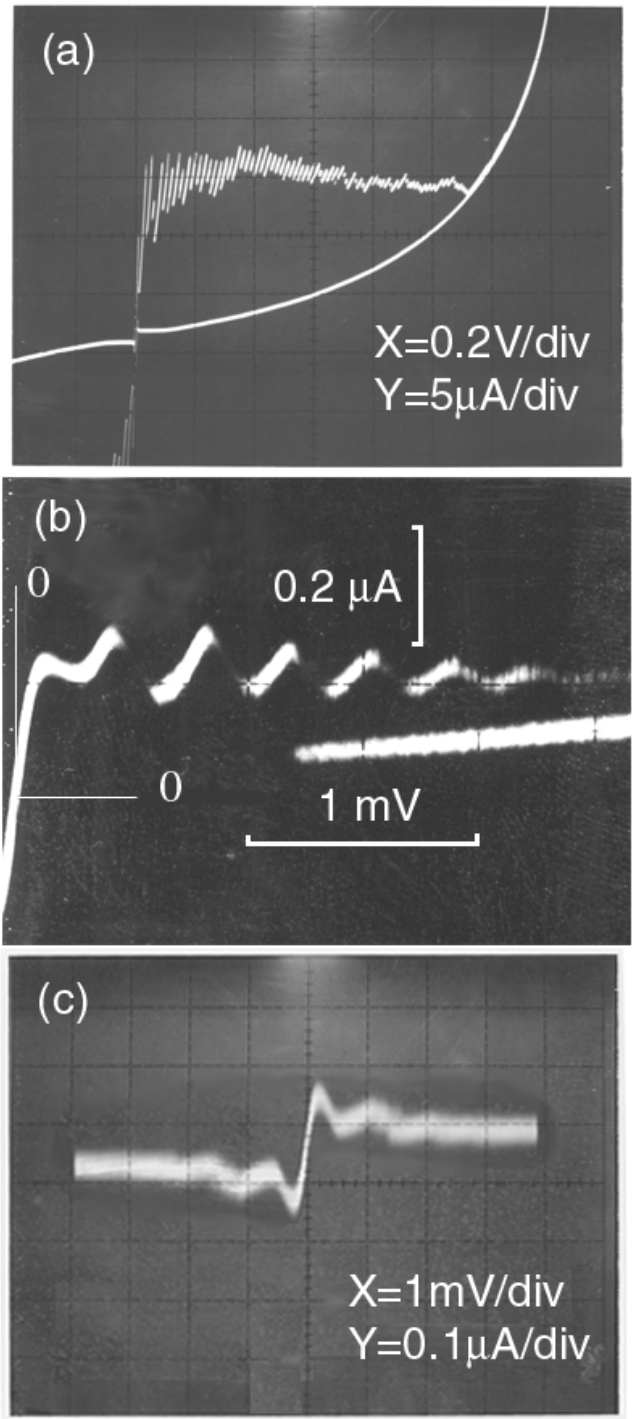

Fig.3. The small current and voltage scale I-V characteristics of the *Bi-2212* stacks:
(a) #2, $S = 2$ μm$^2$; (b) #4, $S = 0.6$ μm$^2$; (c) #6, $S = 0.3$ μm$^2$. T = 4.2 K.

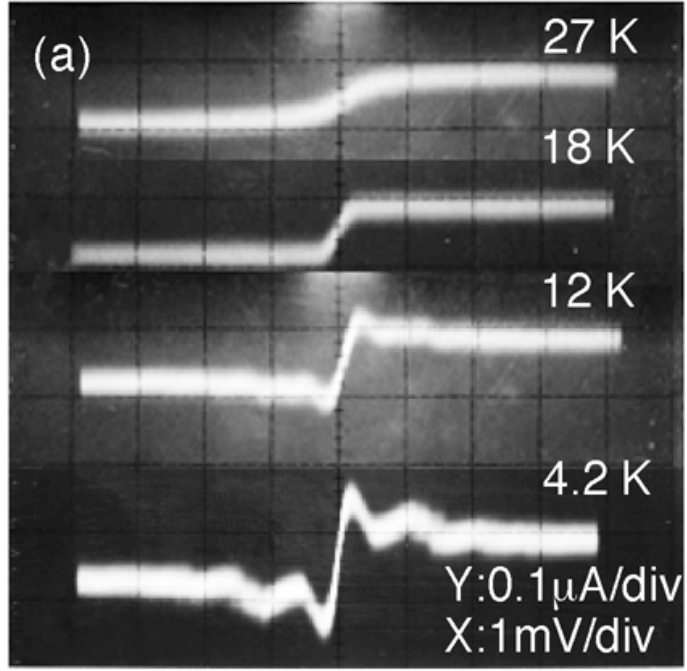

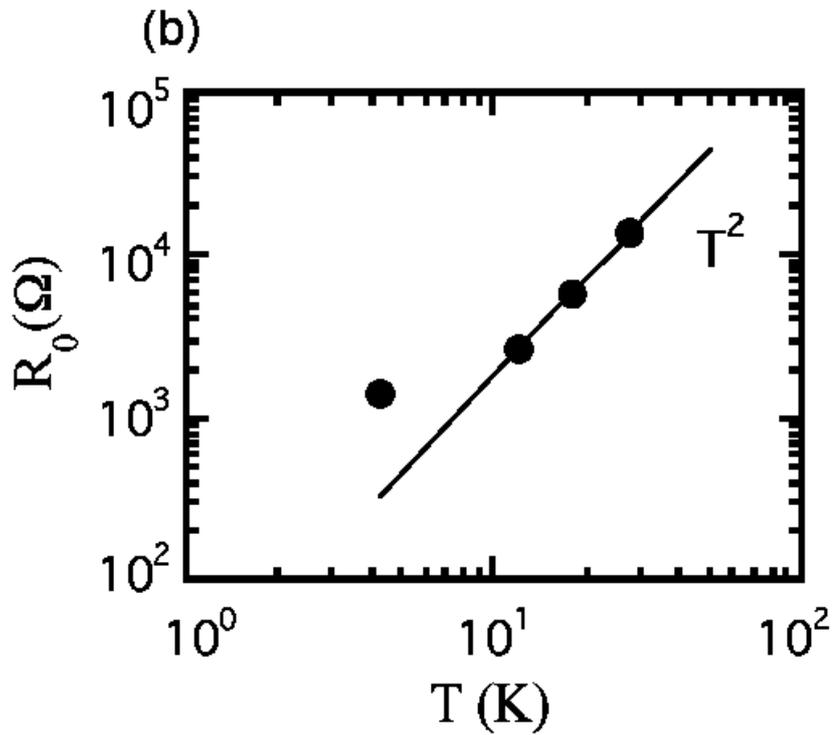

Fig. 4. Temperature evolution of the periodic structure at the *I-V* characteristics (a) and the temperature dependence of the zero bias resistance $R_0$ (b) of the *Bi-2212* stack #6, $S = 0.3$ μm$^2$.

13